\documentclass[twocolumn,prl,superscriptaddress,showpacs,aps,10pt]{revtex4-1}
\newcommand{\pagenumbaa}{1}
\usepackage{graphicx}
\usepackage{color}

\usepackage{units}
\begin{document}


\title{Robust Gapless Surface State against Surface Magnetic Impurities on (Bi$_{0.5}$Sb$_{0.5}$)$_2$Te$_3$ Evidenced by \textit{In Situ} Magnetotransport Measurements}


\author{Liuqi Yu}
\affiliation{Department of Physics, Florida State University, Tallahassee, Florida 32306, USA}

\author{Longqian Hu}
\affiliation{Department of Physics, Florida State University, Tallahassee, Florida 32306, USA}

\author{Jorge L. Barreda}
\affiliation{Department of Physics, Florida State University, Tallahassee, Florida 32306, USA}

\author{Tong Guan}
\affiliation{Beijing National Laboratory for Condensed Matter Physics, Institute of Physics, Chinese Academy of Sciences, Beijing 100190, China}

\author{Xiaoyue He}
\affiliation{Beijing National Laboratory for Condensed Matter Physics, Institute of Physics, Chinese Academy of Sciences, Beijing 100190, China}

\author{Kehui Wu}
\affiliation{Beijing National Laboratory for Condensed Matter Physics, Institute of Physics, Chinese Academy of Sciences, Beijing 100190, China}
\affiliation{Songshan Lake Materials Laboratory, Dongguan, Guangdong 523808, China}

\author{Yongqing Li}
\affiliation{Beijing National Laboratory for Condensed Matter Physics, Institute of Physics, Chinese Academy of Sciences, Beijing 100190, China}
\affiliation{Songshan Lake Materials Laboratory, Dongguan, Guangdong 523808, China}

\author{Peng Xiong}
\email{pxiong@fsu.edu}\affiliation{Department of Physics, Florida State University, Tallahassee, Florida 32306, USA}

\begin{abstract}
 
Despite extensive experimental and theoretical efforts, the important issue of the effects of surface magnetic impurities on the topological surface state of a topological insulator (TI) remains unresolved. We elucidate the effects of Cr impurities on epitaxial thin films of (Bi$_{0.5}$Sb$_{0.5}$)$_{2}$Te$_{3}$: Cr adatoms are incrementally deposited onto the TI held in ultrahigh vacuum at low temperatures, and \textit{in situ} magnetoconductivity and Hall effect measurements are performed at each increment with electrostatic gating. In the experimentally identified surface transport regime, the measured minimum electron density shows a non-monotonic evolution with the Cr density ($n_{\mathrm{Cr}}$): it first increases and then decreases with $n_{\mathrm{Cr}}$. This unusual behavior is ascribed to the dual roles of the Cr as ionized impurities and electron donors, having competing effects of enhancing and decreasing the electronic inhomogeneities in the surface state at low and high $n_{\mathrm{Cr}}$ respectively. The magnetoconductivity is obtained for different $n_{\mathrm{Cr}}$ on one and the same sample, which yields clear evidence that the weak antilocalization effect persists and the surface state remains gapless up to the highest $n_{\mathrm{Cr}}$, contrary to the expectation that the deposited Cr should break the time reversal symmetry and induce a gap opening at the Dirac point. 

\end{abstract}
\maketitle

\setcounter{page}{\pagenumbaa}
\thispagestyle{plain}



Distinct from ordinary band insulators, a three-dimensional (3D) topological insulator (TI) has a topologically protected surface state characterized by Dirac band dispersion and a helical spin texture\cite{Hasan2010,Qi2011,Hsieh2008,Chen2009,Xia2009,Moore2007,Fu2007,Zhang2009}. Protected by time-reversal symmetry (TRS), the surface states (SSs) of TIs are robust against weak nonmagnetic perturbations and may even be immune from Anderson localization. In contrast, the exchange coupling between the localized magnetic moments and surface electrons breaks TRS and is expected to open a gap at the Dirac point\cite{Hasan2010,Qi2011,Liu2009}. The helical gap hosted by the topological SSs has been associated with a number of predictions of exotic quantum phenomena\cite{Qi2009,Nomura2011}. In principle, the most straightforward way for the introduction of SS gap and experimental realization of the predicted effects is via doping of magnetic impurities (MIs). One of the most remarkable examples is the successful realization of the quantum anomalous Hall effect (QAHE) in transition metal-doped TIs, e.g., Cr or V doped into the bulk of (Bi,Sb)$_{2}$Te$_{3}$ films in epitaxial growth\cite{Chang2013,Checkelsky2014, Chang2015}. Ferromagnetic ordering and resulting gap opening are observed in narrow ranges of doping concentrations and QAHE emerges when the Fermi level is tuned into the induced gap. Despite the success, the microscopic mechanism for the MI-induced surface gap remains inconclusive\cite{Chen2010b,Hor2010,Okada2011,He2011,Xu2012,Liu2012,Checkelsky2012,Chang2014,Wray2011,Scholz2012,Valla2012,Honolka2012,Schlenk2013,Sessi2014,Yang2013}. In contrast to bulk doping, both the experimental and theoretical situations with \textit{surface} MI doping of TIs are far more uncertain. Early ARPES experiments\cite{Wray2011} appeared to indicate that deposition of Fe impurities on the surface of Bi$_{2}$Se$_{3}$ induces a gap opening at the Dirac point of the SS. However, more recent ARPES measurements\cite{Scholz2012,Schlenk2013} showed a surprisingly robust SS of Bi$_{2}$Se$_{3}$ without any measurable gap opening at significant amount of Fe impurities on the surface. This is surprising in that in comparison to the insulating bulk, the conducting SS should be more conducive to mediating long-range ferromagnetic ordering. These observations have led to theoretical proposals\cite{Biswas2010,Schaffer2012,Schaffer2015} suggesting that the strong potential from MIs produces electronic states near the Dirac point which could fill the magnetically induced gap. 

\begin{figure}
\includegraphics[width=1\columnwidth]{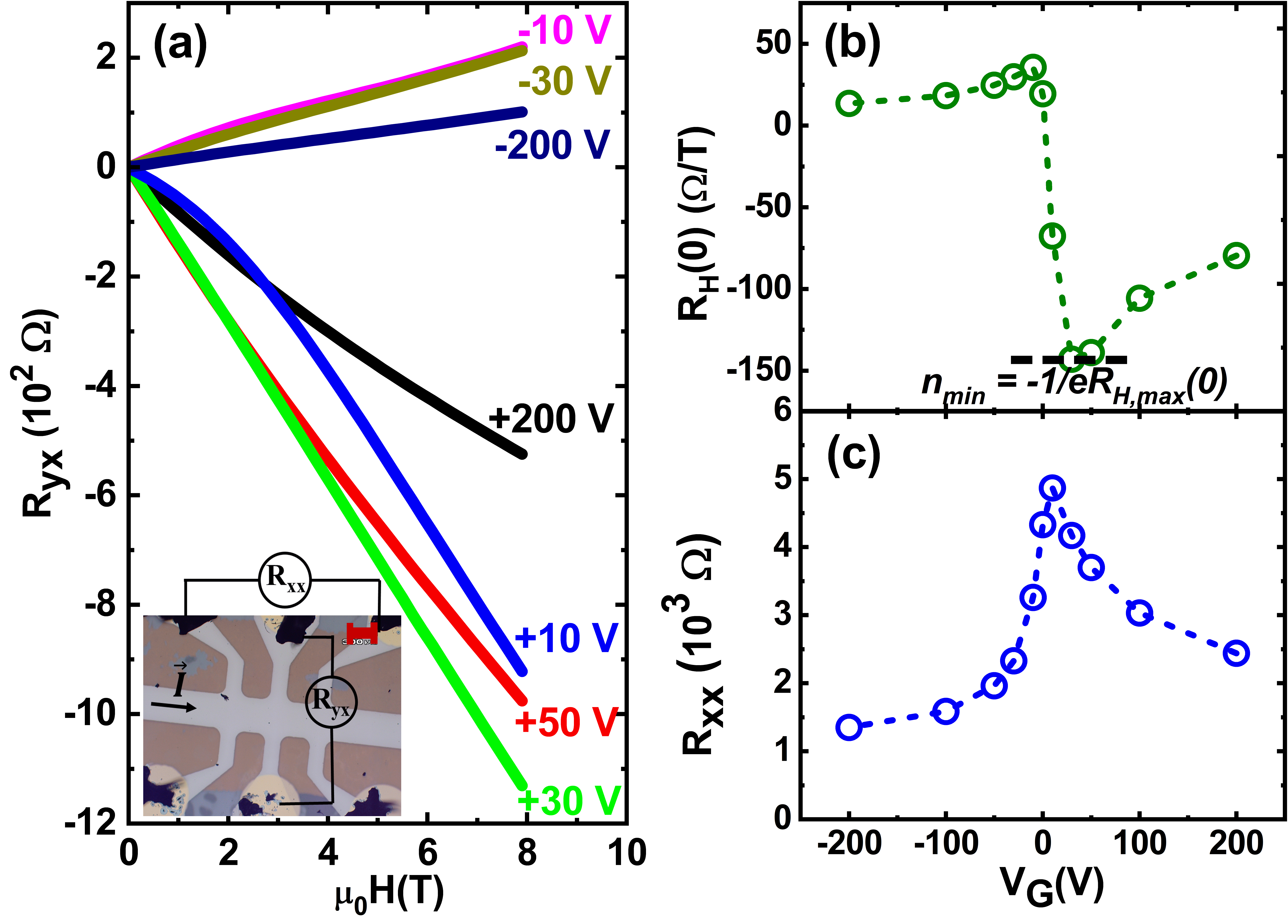}
\caption{Ambipolar field effect measured on a $15\ nm$ thick (Bi$_{0.5}$Sb$_{0.5}$)$_{2}$Te$_{3}$ epitaxial film on SrTiO$_{3}$ (111). The measurement temperature T $= 500\ mK$. (a) Hall resistivity curves at selected $V_{\mathrm{G}}$'s. Inset: optical image of the patterned TI device. The red scale bar measures $300\ \mu m$. (b) Zero-field Hall coefficient, $R_{\mathrm{H}}$(0) and, (c) Corresponding sheet resistance, $R_{\mathrm{xx}}$, with varying $V_{\mathrm{G}}$.}
\label{fig:1}
\end{figure}


\begin{figure*}
\center
\includegraphics[width=0.9\textwidth]{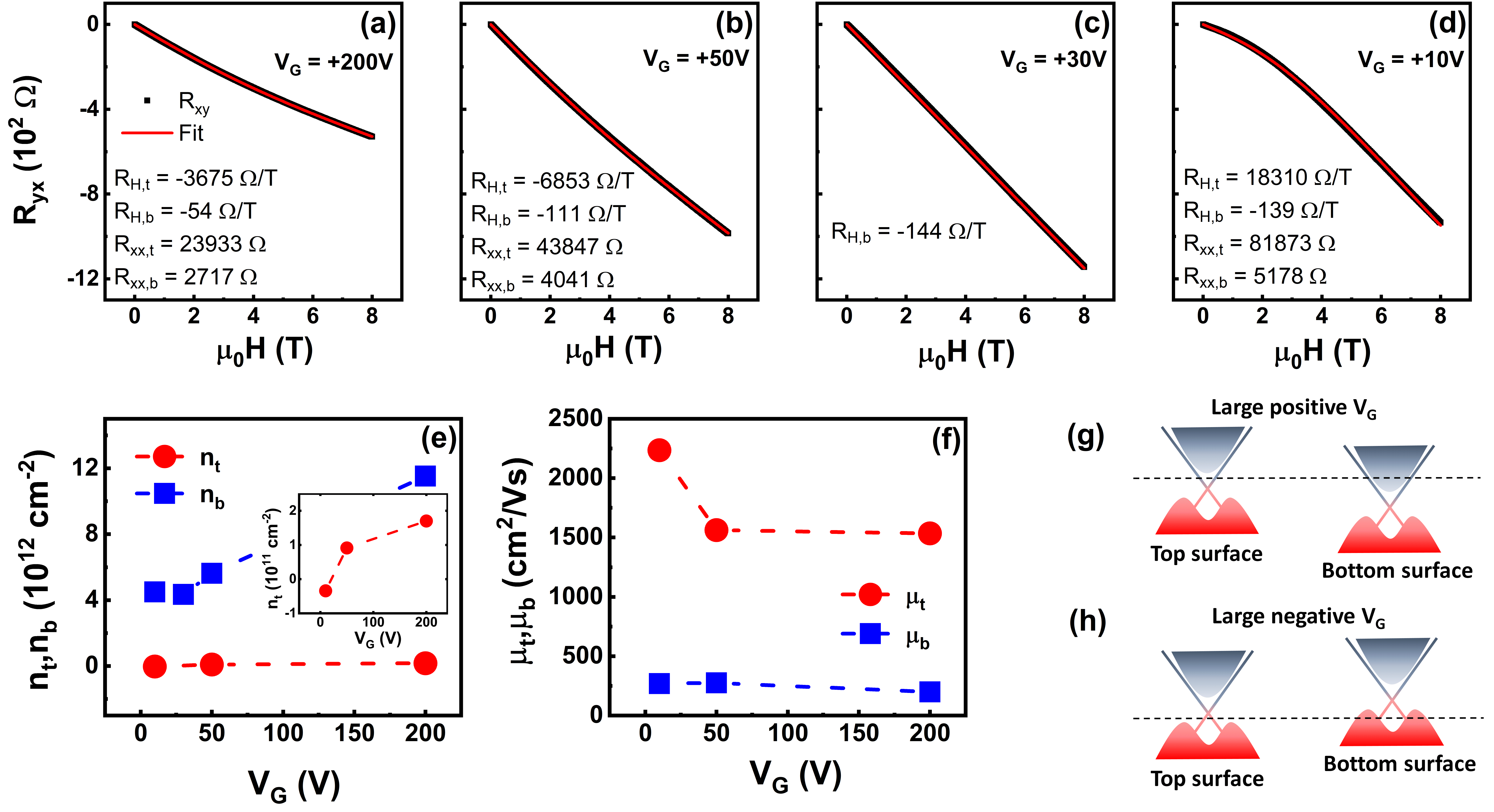}
\caption{(a)-(d) Two-band fittings of the Hall resistivity at $V_{\mathrm{G}}$ = +200 $V$, +50 $V$, +30 $V$ and +10 $V$, respectively. $R_{\mathrm{H,t}}$, $R_{\mathrm{H,b}}$ and $R_{\mathrm{xx,t}}$, $R_{\mathrm{xx,b}}$ denote the Hall coefficients and sheet resistances for the top and bottom SSs. At $V_{\mathrm{G}}$ = +30 $V$, the $R_{\mathrm{yx}}$ curve is essentially linear, which precludes a meaningful two-band fitting. The resultant Hall coefficient from a linear fit is assigned to $R_{\mathrm{H,b}}$ of the bottom surface, which dominates in this transport regime. (e)(f) Carrier densities, $n_{\mathrm{t}}$ and $n_{\mathrm{b}}$, and the corresponding mobilities $\mu_{\mathrm{t}}$ and $\mu_{\mathrm{b}}$ for the top and bottom SS at each $V_{\mathrm{G}}$, respectively. Inset: close-up view of $n_{\mathrm{t}}$ at each $V_{\mathrm{G}}$. (g) (h) Schematic band structures for the top and bottom SS at large positive and negative  $V_{\mathrm{G}}$’s, respectively.}
\label{fig:2}
\end{figure*}


Experimentally, an effective quantitative probe of the magnetic effects of the MIs is magnetotransport measurements of the weak antilocalization (WAL) effect. The WAL, in the form of a negative magnetoconductance (MC) at low fields, can be analyzed in the model of Hikami-Larkin-Nagaoka (HLN)\cite{Hikami1980}, which, in TIs, depends on the spin-flip scattering of the surface electrons as well as their interactions with the bulk\cite{Garate2012,Chen2010a,Chen2011,Steinberg2011}. A straightforward expectation from the theory is that when a TI is doped with MIs, the WAL should be suppressed and may even transition to weak localization\cite{Lu2011}. However, the suppression of WAL was observed only in Bi$_{2}$Te$_{3}$ thin films deposited with Fe impurities\cite{He2011}.There are two main experimental complications. One is the presence of bulk carriers and their interaction with the MIs. These material and device issues may be addressed by means of chemical doping\cite{Chen2009,Ren2010,Zhang2011,Kong2011,Checkelsky2011} and electrostatic gating\cite{Chen2010a,Chen2011,Steinberg2011,He2012,Kim2012,Yang2014}. Another notable source of uncertainty is that different levels of doping are realized in different samples, making it difficult to precisely determine the MI concentration dependence of any measured quantity from large number of separate devices. 

We therefore conduct measurements on (Bi$_{0.5}$Sb$_{0.5}$)$_{2}$Te$_{3}$ thin films which tends to have more insulating bulk than Bi$_{2}$Te$_{3}$ and Bi$_{2}$Se$_{3}$ \cite{Zhang2011,Kong2011,He2012} so as to provide a well isolated top SS for evaluating the effects of deposited MI. As evidenced by both the Hall effect (HE) and MC measurements, the electron transport can be tuned to the surface dominated regime by electrostatic gating. For the MI doping, we employed a technique of \textit{in situ} quench-condensation onto the top surface of a TI held at cryogenic temperatures\cite{Xiong2012,Parker2006,Gardner2011}. Cr adatoms were deposited in controlled increments, and after each deposition a full set of measurements of ambipolar field effect and WAL was performed on \textit{one and the same} sample. The setup therefore eliminates sample-to-sample variation and any air exposure between MI depositions. The results enable us to identify the effects of the Cr impurities on the top SS and their evolution with the Cr density, $n_{\mathrm{Cr}}$. WAL was found to persist up to the highest $n_{\mathrm{Cr}}$, and its analysis suggests absence of any gap opening in the SS. We emphasize that the conclusion is solid even if there was some surface degradation from the initial air exposure, since we relied on the \textit{evolution} of the HE and MC with $n_{\mathrm{Cr}}$ in the same sample.

The TI samples were epitaxial (Bi$_{0.5}$Sb$_{0.5}$)$_{2}$Te$_{3}$ thin films grown on SrTiO$_{3}$ (STO) (111) substrates ($300\ \mu m$ thick) by molecular beam epitaxy\cite{He2012}. The thin films were patterned into a Hall bar by ion milling with a shadow mask. A backgate was thermally evaporated onto the back of the STO substrate (See Supplemental Material, S1, which includes Ref [46]). The Cr deposition is done by resistive heating of a NiCr wire at fixed current. Verification of the deposition of Cr and its possession of a magnetic moment, and calibration of its (relative) density were performed through separate measurements of suppression of the superconducting $T_{\mathrm{C}}$ of ultrathin Pb films\cite{Xiong2012,Parker2006,Gardner2011} (See Supplemental Material, S2, which includes Refs [48,49]). The Cr adatoms thus deposited unambiguously possess magnetic moment and there is effective spin exchange scattering. In the following, the Cr density is indicated by the cumulative deposition time in seconds.

For as-grown (Bi$_{0.5}$Sb$_{0.5}$)$_{2}$Te$_{3}$, the Dirac point of the SS is in close proximity to the top of the bulk valence band\cite{Zhang2011,Kong2011}. Electrostatic gating is used to tune the chemical potential and attain a state of surface-dominated transport. An indication of effective tuning of the chemical potential is the observation of ambipolar field effect, which is evidenced in the Hall effect [Fig.1(a)] for a $15\ nm$ thick (Bi$_{0.5}$Sb$_{0.5}$)$_{2}$Te$_{3}$ thin film [inset of Fig.1(a)] with varying $V_{\mathrm{G}}$. The Hall curves show varying degrees of nonlinearities over the gating range, indicating significant differences in the carrier densities and/or mobilities of the two SSs. In Fig.1(b) and 1(c), we plot the zero-field Hall coefficient, $R_{\mathrm{H}}$(0), and the corresponding sheet resistance, $R_{\mathrm{xx}}$, respectively, as functions of $V_{\mathrm{G}}$. The electron density reaches a minimum value of about $4.4 \times 10^{12}\ cm^{-2}$ at $V_{\mathrm{G}}$ = +30 $V$ (electron density $n = -1/eR_{H}$, where $e$ is the electron charge).

To ascertain whether the chemical potential aligns close to the Dirac point of the top SS, we fit the HE resistivities to the two-band model (See Supplemental Material, S3), as shown in Figs.2(a)-(d). This is applied only to Hall data of positive $V_{\mathrm{G}}$’s; for negative $V_{\mathrm{G}}$’s, satisfactory two-band fitting cannot be obtained, probably due to the presence of bulk conduction. For positive $V_{\mathrm{G}}$’s, the bulk conduction is considered negligible, namely, $1/R_{xx}=1/R_{xx,t}+1/R_{xx,b}$, where $R_{\mathrm{xx}}$ is shown in Fig. 1(c). Figures 2(e) and 2(f) show the resulting carrier densities $n_{\mathrm{t}}$, $n_{\mathrm{b}}$ and mobilities $\mu_{\mathrm{t}}$, $\mu_{\mathrm{b}}$ for the top and bottom SSs. Their $V_{\mathrm{G}}$ dependences are consistent with the chemical potential being raised up and away from the Dirac point by positive $V_{\mathrm{G}}$’s. It should be noted that the Hall resistivity up to 8 $T$ may not have sufficient nonlinearity for a reliable fitting. Measurements of 18 $T$ were performed on another device (See Supplemental Material, S3). The main characteristics are qualitatively similar: $n_{\mathrm{t}}$ is more than an order of magnitude smaller than $n_{\mathrm{b}}$, and its dependence on gating is significantly weaker. The mobility is calculated as $\mu=R_{H}/R_{xx}$ and shown in Fig.2(f). The mobility for the free top surface is almost 10 times higher than that of the bottom surface on the substrate. These observations suggest band structures depicted schematically in Figs.2(g) and 2(h) for large positive and negative $V_{\mathrm{G}}$’s. Due to electric screening, the backgate should have much diminished modulation of the top SS. The chemical potential is located close to the Dirac point of the top SS without any gating, and the surface dominated transport regime can be reached by a positive $V_{\mathrm{G}}$.

\begin{figure}
\includegraphics[width=1\columnwidth]{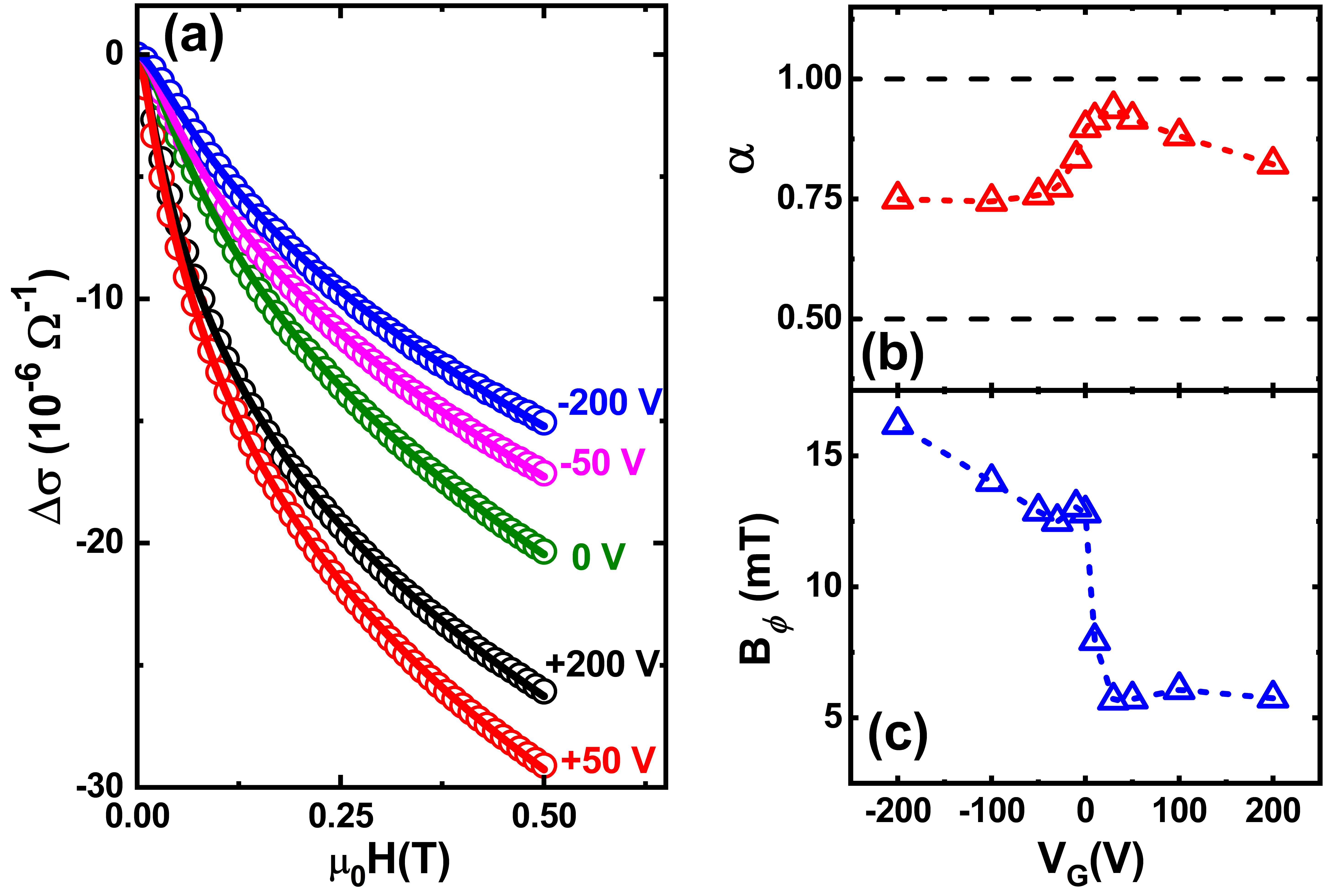}
\caption{(a) Magnetoconductivity for the same sample at selected gate voltages at $T = 500\ mK$. Magnetoresistance at higher magnetic fields is shown in Supplemental Material, S4, which includes Refs [50,51]. The solid lines are fits to the HLN equation. Further details and remarks regarding the fitting can be found in the Supplemental Material, S5. (b) (c) Best-fit values of $\alpha$ and $B_{\mathrm{\phi}}$ from fittings to the HLN equation at each $V_{\mathrm{G}}$.}
\label{fig:3}
\end{figure}

To identify the surface transport regime, a complementary and more rigorous method is to analyze the WAL effect\cite{Chen2010a,Chen2011}, as shown in Fig.3(a) for selected $V_{\mathrm{G}}$’s. Quantitatively, $\Delta\sigma$ can be described by the simplified HLN equation:

\begin{eqnarray}
\Delta\sigma(B) & = & \sigma(B)-\sigma(0) \\
& = & \alpha\frac{e^2}{2\pi^2\hbar}[\ln{\frac{B_\phi}{B}}-\psi(\frac{1}{2}+\frac{B_\phi}{B})]
\label{eq:HLN}
\end{eqnarray}

\noindent
where B is the applied magnetic field, $\hbar$ is Planck’s constant, $\psi$ is the digamma function. $B_\phi$ is defined as the dephasing field. $\alpha$ is a fitting parameter, which can serve as a measure of the channel separation: $\alpha$ is equal to 0.5 for a single 2D channel. It has been established\cite{Chen2010a,Chen2011,Steinberg2011,Bansal2012,Kim2013} that in a TI, the top and bottom SSs can be decoupled by electrostatic gating: with varying $V_{\mathrm{G}}$, $\alpha$ can change from 0.5 to approximately 1, which is interpreted as a signature of the transition from a single transport channel (coupled top and bottom surfaces via a conducting bulk) to two independent channels (insulating bulk and decoupled top and bottom surfaces). As shown in Fig.3(b), the SS dominated transport regime is identified at $V_{\mathrm{G}}$’s where $\alpha$ reaches maximum (closest to 1). Small deviation of $\alpha$ from 1 can be caused by any asymmetry in the top and bottom surfaces even when they are decoupled\cite{Chao2013}, as is evident in Fig.2. The variation of $B_{\mathrm{\phi}}$ with gate voltage is shown in Fig.3(c). $B_{\mathrm{\phi}}$ shows a sharp drop from the \textit{hole} to \textit{electron} transport.

\begin{figure}
\includegraphics[width=1\columnwidth]{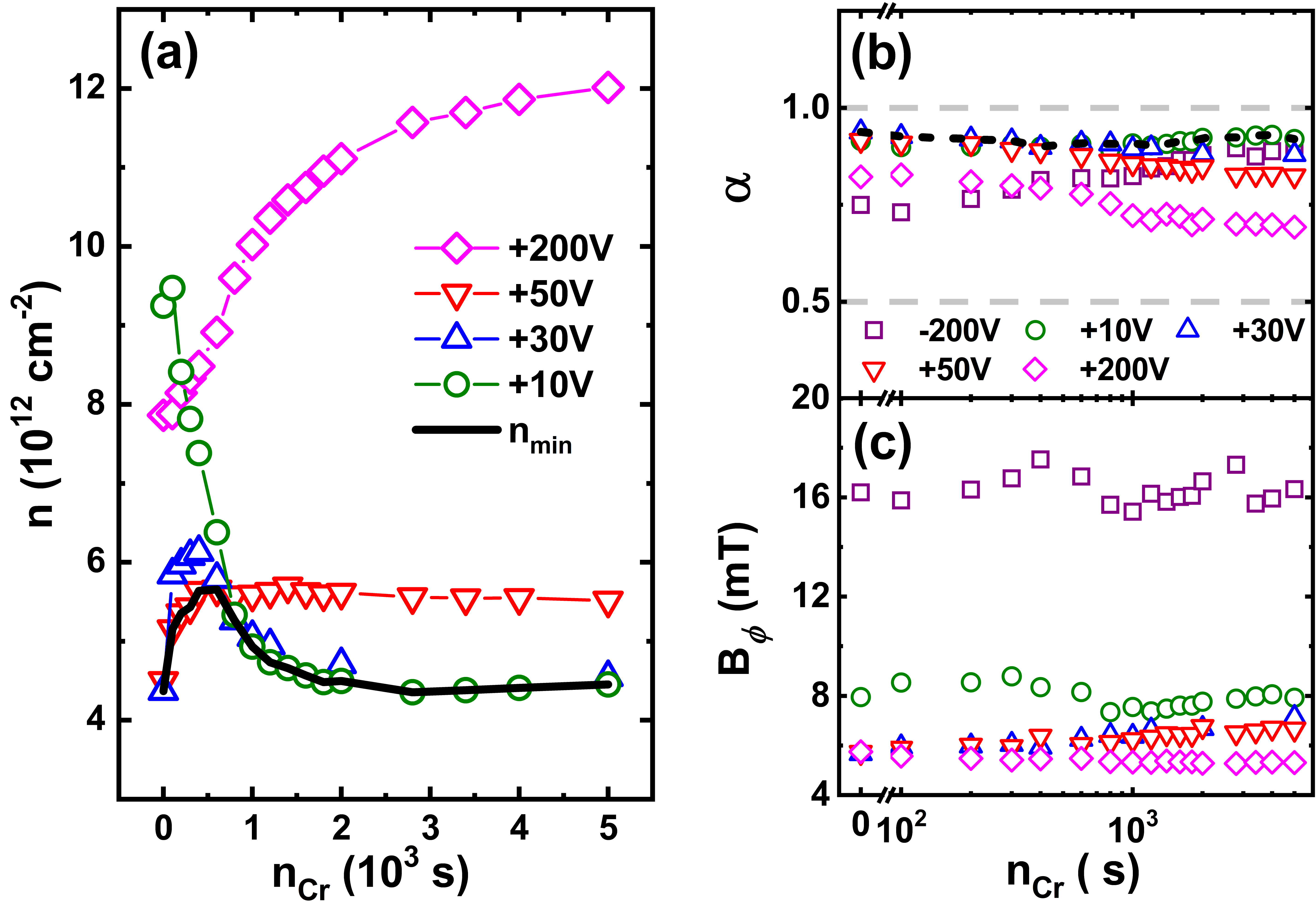}
\caption{(a) Electron densities at selected $V_{\mathrm{G}}$’s with increasing Cr density for the same sample. $V_{\mathrm{G}}$ = +10 $V$, +30 $V$, and +50 $V$ are close to the surface transport regime. $V_{\mathrm{G}}$ = +200 $V$ serves as a reference showing the effect of electron doping by the Cr impurities. The black solid line marks the minimum electron density ($n_{\mathrm{min}}$) identified at each $n_{\mathrm{Cr}}$.(b) $\alpha$ and (c) $B_{\mathrm{\phi}}$ from fittings of the MC data to the HLN equation at selected $V_{\mathrm{G}}$’s as functions of Cr density.}
\label{fig:4}
\end{figure}


Figure 4(a) shows carrier densities of the same sample upon \textit{in situ} incremental Cr deposition at selected $V_{\mathrm{G}}$’s. In the decoupled regime, we assume that the bottom SS is not subject to the influence of the Cr impurities. The change of the overall electron density is then attributed to the top surface. The effect of electron doping of the Cr adatoms on the top surface is apparent at $V_{\mathrm{G}}$ = +200 $V$, where the electron density increases monotonically with $n_{\mathrm{Cr}}$. In the vicinity of the surface transport regime, the evolution of the carrier density is more complex. For each $n_{\mathrm{Cr}}$, we identify a minimum electron density, $n_{\mathrm{min}}$, which is indicated by the black solid line in Fig.4(a). $n_{\mathrm{min}}$ = $-1/eR_{H,max}(0)$, where $R_{\mathrm{H,max}}$(0) is the maximum Hall coefficient at a particular $n_{\mathrm{Cr}}$ in the entire gating range. An example of $R_{\mathrm{H,max}}$(0) at $n_{\mathrm{Cr}}$ = 0 s is indicated by the black dashed line in Fig.1(b). $n_{\mathrm{min}}$ thus determined shows a non-monotonic dependence on the Cr density. $n_{\mathrm{min}}$ thus determined increases initially with $n_{\mathrm{Cr}}$, and then decreases and approaches a constant upon further increase of $n_{\mathrm{Cr}}$. The non-monotonic dependence of $n_{\mathrm{min}}$ on $n_{\mathrm{Cr}}$ is not expected from carrier doping of an electronically homogeneous system, as exemplified by the $V_{\mathrm{G}}$ = +200 $V$ state. Instead, this can be understood as a result of the electronic inhomogeneities in the SS in the form of electron-hole puddles near the Dirac point\cite{Kim2012,Beidenkopf2011,Liao2017}. In the dilute limit, the Cr primarily serve as charged impurities, which increase local potential fluctuations. The increasing ionized impurity scattering can be characterized by increasing $n_{\mathrm{min}}$. This effect of Coulomb scattering has been observed in both graphene\cite{Tan2007,Adam2007} and TI\cite{Kim2012}. At higher Cr densities, the SS becomes electrically more homogeneous, the local potential fluctuations diminish, and consequently $n_{\mathrm{min}}$ decreases. In our experiments, $n_{\mathrm{min}}$ reaches a maximum at $n_{\mathrm{Cr}}$ = 600 $s$, which we surmise is the percolation threshold for a homogeneous Cr-doped top SS. This is consistent with the top SS initially being close to the Dirac point and with a proliferation of electron/hole puddles, similar to the reported results from transport measurements on Bi$_2$Se$_3$ thin films\cite{Kim2012} and STM measurements on Bi$_2$Te$_3$ and Bi$_2$Se$_3$ doped with Ca or Mn impurities\cite{Beidenkopf2011}.

Figure 4(b) and 4(c) shows the results from the analysis of the MC data in the framework of WAL at varying Cr densities, through which we examine the \textit{magnetic} role of the Cr impurities. (See Supplemental Material, S6 for the data at selected $V_{\mathrm{G}}$’s with increasing Cr densities.) In the surface transport regime ($V_{\mathrm{G}}$ = +10 $V$ and +30 $V$), $\alpha$ stays close to 1 (see Fig.4(b)), even at the highest Cr densities. This is strong evidence that both SSs remain gapless in the (large) range of $n_{\mathrm{Cr}}$ studied, since an $\alpha$ of 0.5 is expected with a gapped top surface when the chemical potential is tuned to be positioned within the gap. Interestingly, $\alpha$ at the two opposite ends of the gating range, $V_{\mathrm{G}}$ = +200 $V$ and -200 $V$, show opposite trends with changing $n_{\mathrm{Cr}}$. This is consistent with the effect of electron doping for states as depicted in Fig.2: At $V_{\mathrm{G}}$ = +200 $V$(Fig.2(g)), the electron doping pushes the chemical potential further away from the Dirac point toward the bulk conduction band, leading to a decrease of $\alpha$ due to increased surface-bulk coupling and/or top-bottom surface asymmetry; at the opposite end at $V_{\mathrm{G}}$ = -200 $V$(Fig.2(h)), the electron doping from the deposited Cr raises the chemical potential above the valence band into the surface transport regime for the top surface, leading to an increase of $\alpha$. In contrast, in the surface transport dominated regime ($V_{\mathrm{G}}$ = +10 $V$ and +30 $V$), $\alpha$ stays close to 1 and shows little variation with $n_{\mathrm{Cr}}$. This observation ($\alpha$ $\approx$ 1) in the clearly identified surface transport states provides compelling evidence that WAL persists up to the highest $n_{\mathrm{Cr}}$.

Figure 4(c) shows the resulting dephasing field $B_{\mathrm{\phi}}$ from the same fittings. With significant electronic inhomogeneity or multiple channels of transport, as exemplified by the case of $V_{\mathrm{G}}$ = -200 $V$, the $B_{\mathrm{\phi}}$ thus determined shows pronounced scatter and may not even be physically meaningful. A more interesting case is that of $V_{\mathrm{G}}$ = +10 $V$, where $B_{\mathrm{\phi}}$ initially shows large fluctuations but becomes well-defined above $n_{\mathrm{Cr}}$ = 600 $s$. This is in excellent agreement with the picture of the evolution of the electronic states inferred from $n_{\mathrm{min}}$ in Fig.4(a), namely, increasing $n_{\mathrm{Cr}}$ drives the top SS from an inhomogeneous state characterized by electron-hole puddles to a state of uniform electron transport.

Finally, we comment on the origin and implications of the robustness of the gapless SS against surface paramagnetic impurities. The persistence of WAL and absence of a transition from WAL to WL indicates no gap opening in the SS despite significant amount of MIs. Long-range ferromagnetic order is often considered necessary for a gapped SS\cite{Liu2009}. Unlike in the insulating bulk where the ferromagnetic ordering of the magnetic dopants needs a mechanism such as Van Vleck\cite{Chang2013,Yu2010}, in the conducting SS, the long Fermi wavelength (as long as tens of $nm$’s) of the surface electrons implies that the Ruderman-Kittel-Kasuya-Yosida  interaction could induce ferromagnetism even in very dilute limit of surface magnetic dopants\cite{Liu2009}. However, our experiments suggest the absence of any long-range ferromagnetic order in the top SS in a broad range of surface magnetic dopant densities. 

In summary, using a unique experimental scheme combining incremental low-temperature deposition of magnetic impurities and in situ magnetotransport measurements on one and the same TI sample, we have obtained a detailed close look at the evolution of the topological SS with increasing MI density. The insight gleaned from these experiments would have been extremely difficult, if not impossible, from conventional experiments of measuring many samples at different doping levels. The roles of the Cr as electron donors, ionized impurities, and magnetic scatters are identified. Most notably, the Cr impurities are found to have very limited effect on the SS, as evidenced by the remarkable insensitivity of the WAL effect to $n_{\mathrm{Cr}}$ in a well isolated topological SS.

\begin{acknowledgments}
P. X. acknowledges the support by the DARPA TEE Program (Cooperative Agreement \#D18AC00010) and NSF grant DMR-1905843. The work at IOP is supported by National Natural Science Foundation of China (Project No. 61425015),  National Key Research and Development Program (Project No. 2016YFA0300600), and the Strategic Priority Research Program of Chinese Academy of Sciences (Project No. XDB28000000).
\end{acknowledgments}



\end{document}



\title{SUPPLEMENTAL MATERIAL \\
Robust Gapless Surface State against Surface Magnetic Impurities on (Bi$_{0.5}$Sb$_{0.5}$)$_{2}$Te$_{3}$ Evidenced by \textit{In Situ} Magnetotransport Measurements}


\author{Liuqi Yu}
\affiliation{Department of Physics, Florida State University, Tallahassee, Florida 32306, USA}

\author{Longqian Hu}
\affiliation{Department of Physics, Florida State University, Tallahassee, Florida 32306, USA}

\author{Jorge L. Barreda}
\affiliation{Department of Physics, Florida State University, Tallahassee, Florida 32306, USA}

\author{Tong Guan}
\affiliation{Beijing National Laboratory for Condensed Matter Physics, Institute of Physics, Chinese Academy of Sciences, Beijing 100190, China}

\author{Xiaoyue He}
\affiliation{Beijing National Laboratory for Condensed Matter Physics, Institute of Physics, Chinese Academy of Sciences, Beijing 100190, China}

\author{Kehui Wu}
\affiliation{Beijing National Laboratory for Condensed Matter Physics, Institute of Physics, Chinese Academy of Sciences, Beijing 100190, China}
\affiliation{Songshan Lake Materials Laboratory, Dongguan, Guangdong 523808, China}

\author{Yongqing Li}
\affiliation{Beijing National Laboratory for Condensed Matter Physics, Institute of Physics, Chinese Academy of Sciences, Beijing 100190, China}
\affiliation{Songshan Lake Materials Laboratory, Dongguan, Guangdong 523808, China}

\author{Peng Xiong}
\email{pxiong@fsu.edu}\affiliation{Department of Physics, Florida State University, Tallahassee, Florida 32306, USA}

\maketitle
\setcounter{page}{\pagenumbaa}
\thispagestyle{plain}

\section{Experimental setup}

\begin{figure}[htbp]
	\centering
		\includegraphics[width=0.6\textwidth]{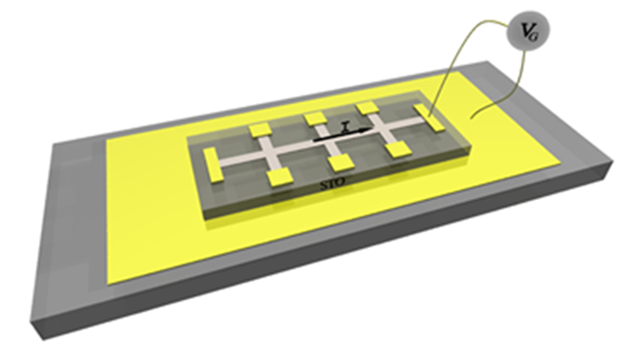}
	\caption{
	A schematic diagram of a patterned device.}
	\label{fig:FigS1}
\end{figure}

Topological insulator (TI) (Bi$_{0.5}$Sb$_{0.5}$)$_{2}$Te$_{3}$ thin film samples used in this work was vacuum sealed in a quartz tube upon removal from the MBE growth chamber until device fabrication. In order to minimize surface contamination in the fabrication process, the sample was patterned into a Hall bar by Ar ion milling with a shadow mask, instead of conventional photolithography and wet etching. The sample temperature during the ion milling was closely monitored and kept below 35$^{\circ}$C. Given that the bombardment of SrTiO$_3$ (STO) by energetic ions could induce oxygen defects thus possibly lead to a conducting surface\cite{Henrich1978}, a control experiment was conducted to completely etch away one TI thin film under similar ion milling conditions, and no measurable conductance on the STO substrate was observed from room temperature down to 380 $mK$. This excludes any possible parallel conduction from the etched areas on STO substrate in a patterned sample. A gold thin film was thermally evaporated onto the back of the STO substrate to serve as a back gate. The MR and Hall effect (HE) measurements were carried out in a customized dilution refrigerator at a fixed sample temperature of 500 $mK$. It is noted that a hysteresis appears in both MR and Hall measurements when cycling the gate voltages, probably due to the charged impurity defects. Thus, to maintain consistency, the gate voltage V$_G$ was applied in the same routine of decreasing from +200 V to -200 V after each Cr deposition. Four different samples from 15 $nm$ to 20 $nm$ thick were studied, and the results are qualitatively consistent; the data obtained from a 15 $nm$ thick sample is presented and discussed in the main text.

\clearpage
\section{Calibration of Cr deposition rate and density}

\begin{figure}[htbp]
	\centering
		\includegraphics[width=1\textwidth]{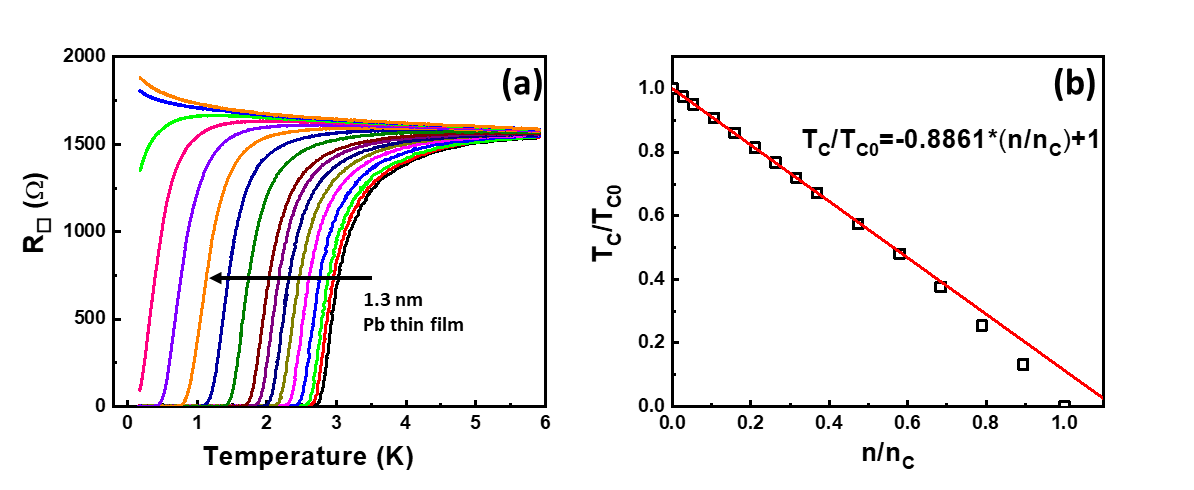}
	\caption{
	(a) Sheet resistance of a Pb thin film (t = 1.3 $nm$, T$_{C0}$ = 3.0 $K$) as a function of temperature at different MI densities. It undergoes a MI-tuned superconductor-insulator transition. (b) Normalized critical temperature, T$_C$/T$_{C0}$, as a function of relative MI density, n/n$_C$.}
	\label{fig:FigS2}
\end{figure}

In our experiment, the Cr can be deposited in extremely small increments. The deposition rate is below the resolution of the quartz crystal monitor, and an alternative way of calibrating the MI density is needed. Paramagnetic impurities introduce pair-breaking of superconductivity\cite{Parker2006}, we thus calibrate the Cr density through measurements of suppression of the superconducting T$_C$ of \textit{in situ} grown ultrathin Pb films\cite{Parker2006,Gardner2011}.
The calibration was carried out in the same modified dilution refrigerator on a sample held below liquid $^4$He temperature. A layer of Sb ($\sim$1 $nm$) was evaporated first. Followed by an ultrathin Pb layer of 1.3 $nm$. As shown in Fig.S2, the thin film has a T$_C$ of approximately 3.0 $K$. The Cr was deposited with a constant DC current of 1.2 $A$ sent through a 0.01$''$ diameter nichrome (NiCr) wire, resulting in Cr sublimation with no detectable Ni. It has been shown that although the absolute values of the Cr density determined this way is unknown, the relative (normalized) values are highly precise [Fig.S2(b)] as measured by the cumulative deposition time\cite{Parker2006}. In Fig.S2(a), it is demonstrated the T$_C$ decreases with increasing MI density and the thin film can be driven from the superconducting phase to the insulating phase by Cr deposition. The same calibrated filament was then used for experiments on the TI sample. 

\clearpage
\section{Two-band fittings of the Hall resistivity}

\begin{figure}[htbp]
	\centering
		\includegraphics[width=1\textwidth]{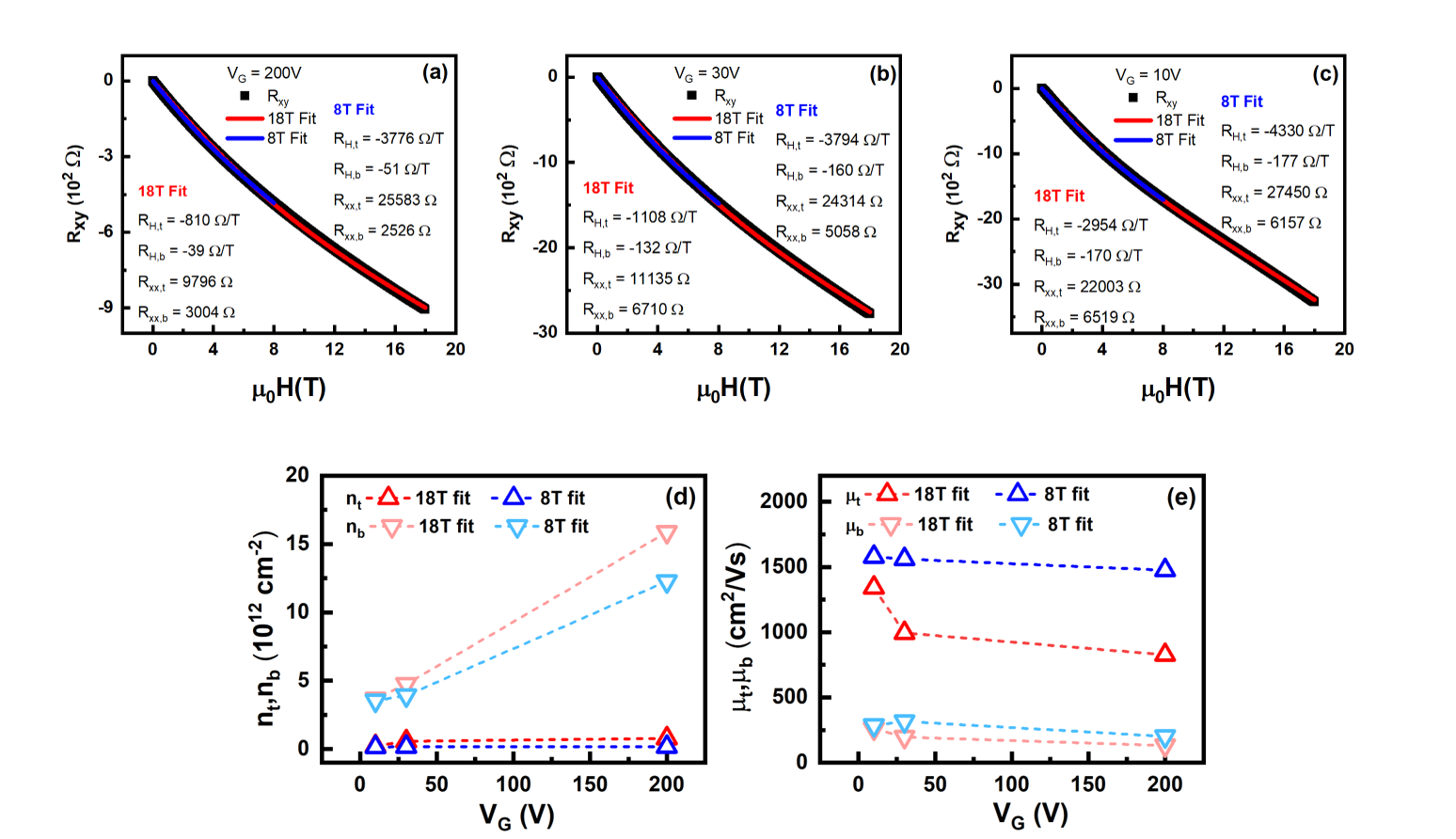}
	\caption{
	(a)-(c)Two-band fittings of the Hall resistivity at selected V$_G$’s up to 18 $T$ measured on a 15 $nm$ thick (Bi$_{0.5}$Sb$_{0.5}$)$_{2}$Te$_{3}$ epitaxial film on SrTiO$_3$ (111) at temperature T = 20 $mK$. Fitting results for field ranges of 18 $T$ and 8 $T$ are shown and compared. (d)(e) Carrier densities, n$_t$ and n$_b$, and the corresponding mobilities $\mu_{t}$ and $\mu_{b}$ for the top and bottom surface states at each V$_G$, respectively. Results from fittings for field ranges of 18 $T$ and 8 $T$ are shown for comparison.}
	\label{fig:FigS3}
\end{figure}

The nonlinearity in the Hall traces (See Fig.1(a) in the main text) suggests more than one conducting channel contribute to electronic transport. The Hall effect characteristic parameters, i.e. carrier densities and mobilities, can be obtained by fitting to the two-band model. The Hall resistivity can be expressed as

\begin{equation}
R_{xy} = \frac{(R_{H,t}R_{xx,b}^2+R_{H,b}R_{xx,t}^2)B+R_{H,t}R_{H,b}(R_{H,t}+R_{H,b})B^3}{(R_{xx,t}+R_{xx,b})^2+(R_{H,t}+R_{H,b})^2B^2}
\end{equation}

\noindent
where R$_{H,t}$, R$_{H,b}$, R$_{xx,t}$, R$_{xx,b}$ correspond to the Hall coefficients and sheet resistances of the top and bottom surfaces, respectively. One can reduce the free parameters for the fitting by one based on the total sheet resistance, $R_{xx}(0)=R_{xx,t}R_{xx,b}/(R_{xx,t}+R_{xx,b})$ at B = 0 $T$, obtained from the corresponding magneto-conductivity measurements. The carrier densities, n$_t$ and n$_b$, in Fig.2(e) are calculated by n$_{t(b)}$ = 1/eR$_{H,t(b)}$, where $e$ is the electron charge. The mobilities, $\mu_{t}$ and $\mu_{b}$, in Fig.2(f) are calculated by $\mu_{t(b)}$=R$_{H,t(b)}$/R$_{xx,t(b)}$. 
Figure S3 show Hall effect results measured on another sample grown and fabricated under the same conditions in magnetic fields up to 18 $T$. Two sets of fittings were conducted using data up to 18 $T$ and 8 $T$, as shown in Figs.S3(a)-(c). The 18 $T$ fittings yield higher carrier densities but lower mobilities for the top surface, in comparison to the 8 $T$ fittings. In both cases, as shown in Fig.S3(e) and S3(f), the top surface has much lower carrier density but higher mobility than the bottom surface, and the top surface shows much weaker response to the back gating.   

\clearpage
\section{High field magnetoresistance}

\begin{figure}[htbp]
	\centering
		\includegraphics[width=1\textwidth]{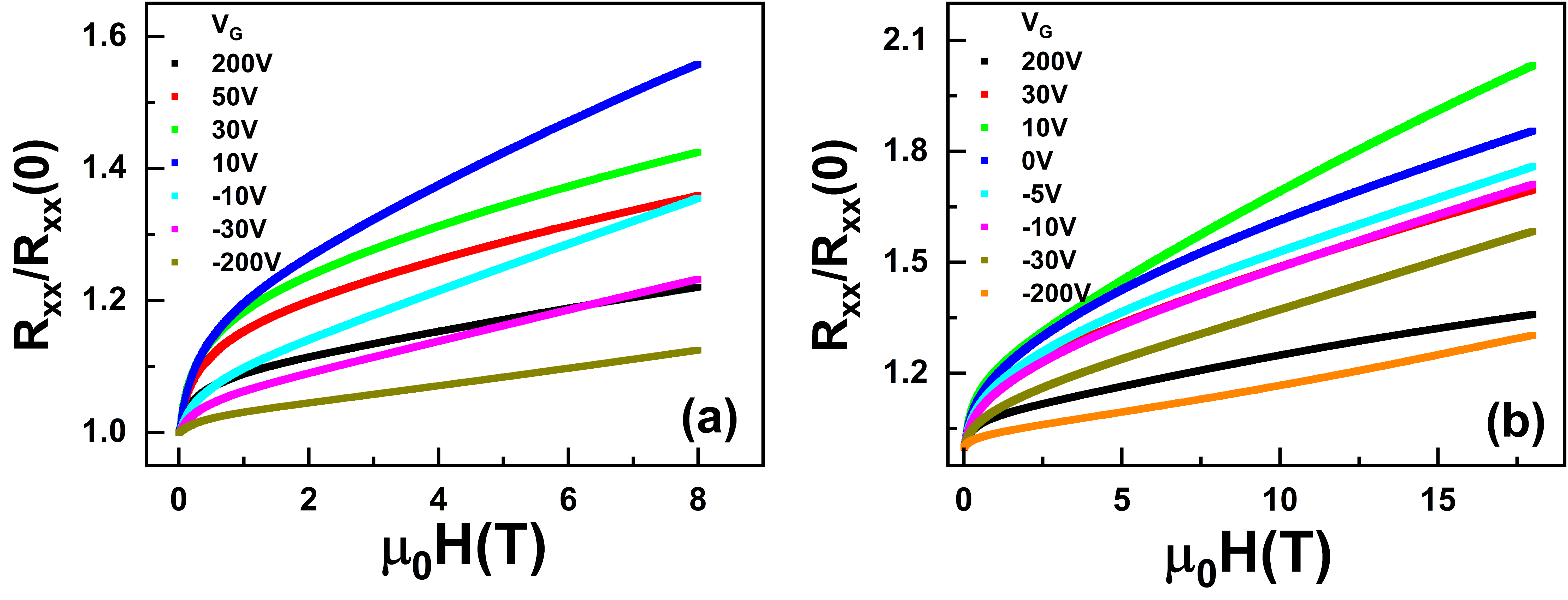}
	\caption{
	(a) MR up to 8 $T$ for the sample discussed in the main manuscript at selected gate voltages at $T = 500\ mK$ (b) MR up to 18 $T$ for a separate sample at selected gate voltages. The measurement temperature $T = 20\ mK$.}
	\label{fig:FigS4}
\end{figure}

The magnetoresistance (MR) at various V$_G$’s up to 8 $T$ is presented in Fig. S4(a) for the sample discussed in the main manuscript. On a separate sample, MR up to 18 $T$ was obtained and shown in Fig. S4(b). On both samples, no quantum oscillations were observed. More interestingly, linear MR appears for a wide range of V$_G$’s, and shows no signs of saturation up to the highest applied magnetic fields. A previous study \cite{Guan2015} indicates the linear MR in (Bi$_{0.5}$Sb$_{0.5}$)$_{2}$Te$_{3}$ most likely originates from electronic inhomogeneities\cite{Parish2003}.  

\clearpage
\section{HLN fittings to weak antilocalization}

\begin{figure}[htbp]
	\centering
		\includegraphics[width=1\textwidth]{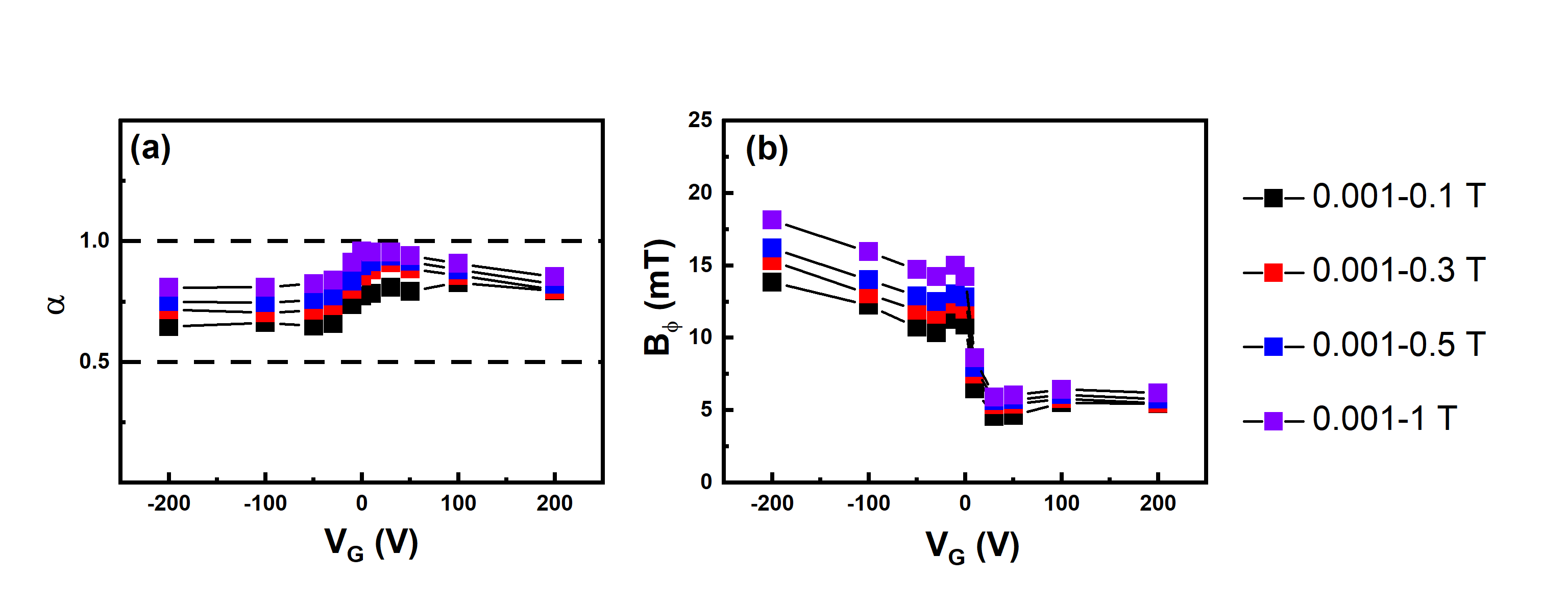}
	\caption{
	Best-fit values of (a) $\alpha$ and (b) B$_\phi$ to the HLN equation with MC data of different magnetic field ranges.}
	\label{fig:FigS5}
\end{figure}

The positive correction to the magnetoconductivity due to WAL is valid in small fields determined by the mean free path l$_m$, the magnetic dephasing length l$_B$, and spin-orbit scattering length l$_{SP}$. For example, the electron phase coherence is lost at high magnetic fields, when l$_B$ $>$ l$_m$. l$_m$ was estimated to be between 5 $nm$ and 15 $nm$ at different V$_G$’s. Therefore, l$_B=\sqrt{\hbar/eB}$, yields an upper field limit of about several tesla. To verify the validity of the fitting, we perform HLN fittings for varying magnetic field ranges, and the resulting best-fit parameters are shown in Fig.S5. Both $\alpha$ and B$_\phi$ show similar trend with different fitting field ranges and are qualitatively consistent. The slight variation probably results from some background magnetoconductivity. In the main text, the HLN fittings are carried out for data in the field range from 0.001 $T$ to 0.5 $T$.

\clearpage

\section{Persistence of WAL effect}

\begin{figure}[htbp]
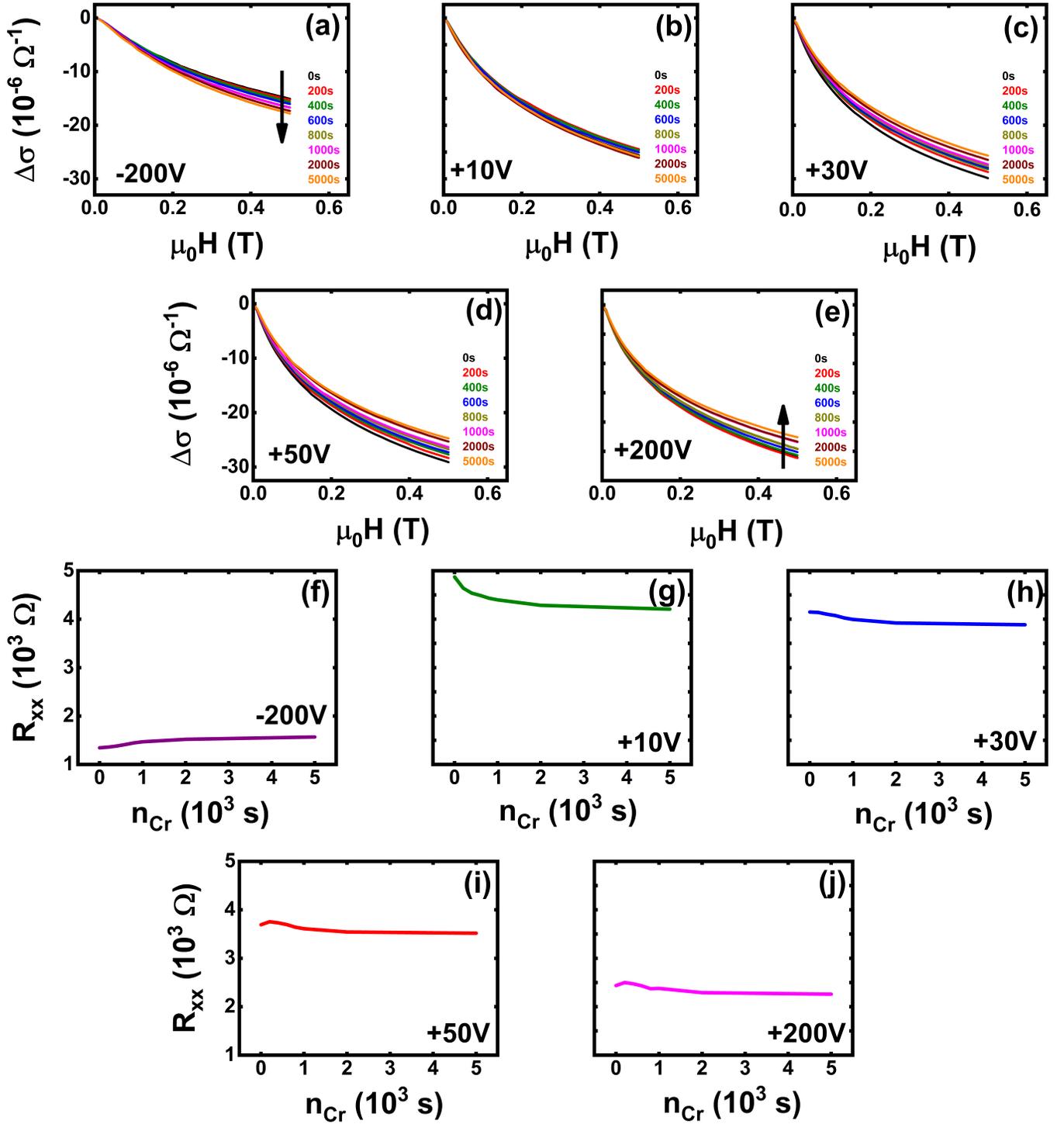

	\centering
	 \includegraphics[width=1\textwidth]{Figure_S6_1.png}
	 \includegraphics[width=1\textwidth]{Figure_S6_2.png}
	\caption{
	(a)-(e) Magnetoconductivities at selected V$_G$’s with increasing Cr densities. (f)-(j) Sheet resistance R$_{xx}$ measured at zero magnetic field at selected at selected V$_G$’s with increasing Cr densities.}
	\label{fig:FigS6}
\end{figure}

The WAL effect persists at all measured V$_G$’s and n$_{Cr}$'s. As indicated by the arrows in Fig. S6(a) and (e), at V$_G$ = +200 $V$ and -200 $V$, the WAL effect shows opposite trends with increasing n$_{Cr}$. This is consistent with the observations of $\alpha$, as presented in Fig. 4(b). As shown in Figs. S6(f) – (j), the deposition of Cr causes noticeable changes in the measured (zero-field) sheet resistance at low n$_{Cr}$ for different gate voltages, which becomes flat at higher n$_{Cr}$. The results suggest that the Cr introduces scattering, but does not form continuous conducting path even at the highest n$_{Cr}$.

\clearpage

